\documentclass[a4paper,11pt]{article}
\pdfoutput=1 

\usepackage{jheppub} 

\usepackage[T1]{fontenc} 

\newcommand{\be}{\begin{equation}}
\newcommand{\ee}{\end{equation}}
\newcommand{\bea}{\begin{eqnarray}}  
\newcommand{\eea}{\end{eqnarray}}  
\usepackage{amsmath}

\DeclareMathOperator{\Tr}{Tr}

\title{\boldmath On the impact of the LHC Run2 data on general Composite Higgs scenarios}
\author[a]{Charanjit K. Khosa}
\author[b,c]{ and Veronica Sanz}

\affiliation[a]{Dipartimento di Fisica, Universit\`a di Genova and INFN, Sezione di Genova,\\ Via Dodecaneso 33, 16146, Italy}
\affiliation[b]{Department of Physics and Astronomy, University of Sussex, \\ BN1 9QH Brighton, UK}
\affiliation[c]{Instituto de F\'isica Corpuscular (IFIC), Universidad de Valencia-CSIC, \\ E-46980, Valencia, Spain}

\emailAdd{ckaur@ge.infn.it}
\emailAdd{v.sanz@sussex.ac.uk }

\abstract{
We study the the impact of Run2 LHC  data on general Composite Higgs scenarios, where non-linear effects, mixing with additional scalars and new fermionic degrees of freedom could simultaneously contribute to the modification of  Higgs properties. We obtain new experimental limits on the scale of compositeness, the mixing with singlets and doublets with the Higgs, and the mass and mixing angle of top-partners. We also show that for scenarios where new fermionic degrees of freedom are involved in electroweak symmetry breaking, there is an interesting interplay among Higgs coupling measurements, boosted Higgs properties, SMEFT global analyses, and direct searches for single- and double-production of vector-like quarks.}

\begin{document} 
\maketitle
\flushbottom

\section{Introduction}
The true origin of the Higgs mechanism is still an open question in Particle Physics, despite the discovery of its key element, the Higgs particle~\cite{atlasHiggs,cmsHiggs}, and the observation of the SM-like nature of its couplings to massive particles~\cite{Higgscouplings}. 

The main reason to doubt a purely SM Higgs sector can be traced back its quantum behaviour, a problem often expressed in the context of {\it naturalness}: how can a fundamental scalar be so light, yet so sensitive to UV effects. And this is not the only suspicious aspect of the SM Higgs. The hierarchy among its Yukawa couplings, or its inability to produce enough baryon asymmetry during the electroweak phase transition also add to the unsatisfactory aspects of the SM Higgs sector.

However, these shortcomings also open opportunities for new physics. The Higgs, due to its scalar nature, could connect to other sectors via mixing with scalars, participate in phase transitions and inflation. Among the theories bringing the Higgs into a new framework, one particularly appealing proposal is the concept of compositeness, leading to Composite Higgs Models (CHMs). In CHMs, the Higgs is not a truly fundamental particle but a bound state of other, more fundamental, particles. 

Primitive proposals for a Composite Higgs~\cite{CHMoriginal} were replaced by more realistic realisations based on the idea that a Higgs is not just any composite state from strong dynamics, but a pseudo-Goldstone boson~\cite{HiggspGBori}.
Pseudo-Goldstone bosons appear when approximate global symmetries are spontaneously broken. The size of the symmetry and what it breaks down to determines the amount of scalar degrees of freedom one will have in CHMs. The partial gauging of some of these global symmetries sets what kinds of quantum numbers the scalars have. The minimal framework to achieve successful Electroweak Symmetry Breaking (EWSB), i.e. leading to a Higgs doublet of $SU(2)_L$ with a $m_h\sim v$, was found in Ref.~\cite{CHMoriginal}. In this setup, the Higgs originated from the breaking $SO(5)\to SO(4)$, and its radiative potential had to be supplemented with a new vector-like fermion with the same quantum numbers as the top, called {\it top-partner} and denoted by $T$.  This minimal set-up and its phenomenological consequences has been thoroughly explored in the literature, see e.g. Ref.~\cite{Furlan} for a review. 

Typically one would assume the main manifestation of the composite nature of the Higgs would be the non-linear origin of its couplings, leading to very specific types of Higgs coupling deviations. Yet Composite Higgs scenarios could exhibit a richer phenomenology, such as the presence of new scalars or fermions which would also modify the Higgs properties. The focus of this paper is the study of the rich patterns which arise in general Composite Higgs scenarios and what the Run2 LHC data can tell us about them. The data we will use for this analysis is summarised in Table~\ref{ExpData}.
 
\begin{table}[t]
\begin{center}
\begin{tabular}{|c|c|}
\hline
Measurement & Reference \\
\hline
{\bf Higgs Measurements} &\\
Run 1 ATLAS and CMS combined Higgs signal strength measurements & \cite{Khachatryan:2016vau} \\
Run 2 ATLAS Higgs signal strength measurements   & \cite{atlasrecent,atlasvhww}  \\
Run 2 CMS Higgs signal strength measurements    &  \cite{cmscombo,cmshtoggnew,cmshtobb,cmshtotautau,cmstthml,cmsmumu}  \\
ATLAS Higgs+jet differential measurement &  \cite{ATLASdiphotondiff}  \\
\hline\hline
{\bf LHC direct searches for VLQ}& \\
ATLAS T $\rightarrow$ Wb channel & \cite{singleVLQsatlaswb} \\
ATLAS T $\rightarrow$ top+MET channel & \cite{AtlastopMET}\\
CMS T $\rightarrow$ Wb channel & \cite{CMSsingleVLQbw} \\
CMS T $\rightarrow$ Zt final state & \cite{CMSsingleVLQZt}\\
\hline
\end{tabular}
\end{center}
\caption{Experimental measurements considered in this work.}
\label{ExpData}
\end{table}

The paper is organised as follows. In section \ref{sec:patterns}, we describe the different patterns arising in Composite Higgs scenarios: non-linear effects (Sec.~\ref{pattern1}), extended scalar sectors (Sec.~\ref{pattern2}) and new fermionic degrees of freedom (Sec.~\ref{pattern3}).  In these section we examine the impact that Higgs measurements and direct top-partner searches  have on these patterns and in combinations of them, e.g. how non-linearities and mixing with new scalars do contribute on the same direction to reduce Higgs couplings to massive particles. Hence, as one switches on more than one of these effects, each individual limit becomes stronger. The last Section~\ref{conclusions} is devoted to conclusions.

\section{Patterns of Composite Higgs scenarios}~\label{sec:patterns}

\begin{figure}[t!]
    \centering
    \includegraphics[scale=0.35]{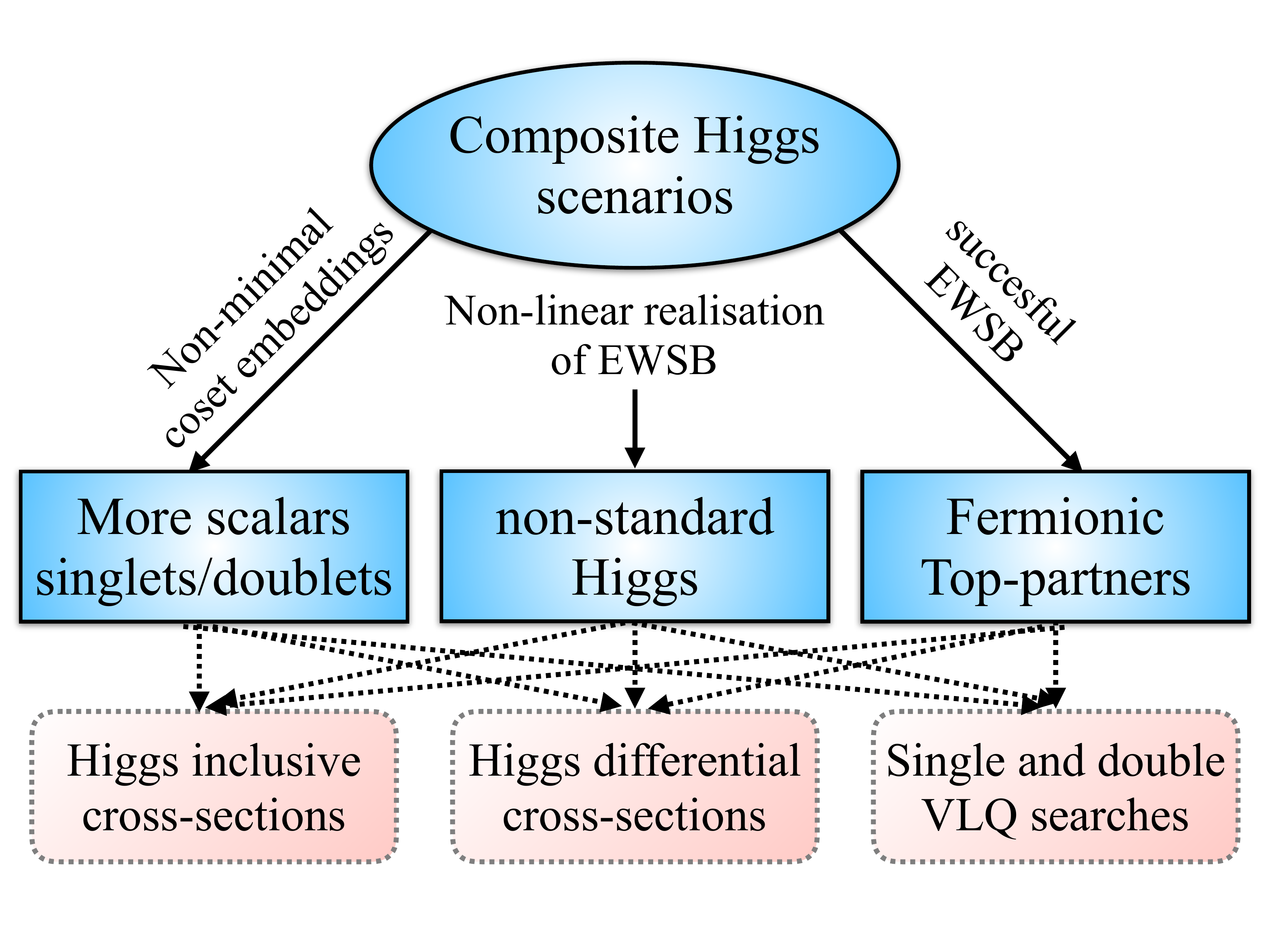}
    \caption{Different realisations of composite Higgs scenario and the relevant experimental measurements.}
    \label{fig:mindmap}
\end{figure}

The idea of the Higgs as a composite state could be realised in many ways. Considering  theoretical constraints and precision measurements  substantially reduces the possible scenarios,  yet there are still many possibilities to contemplate. 
Nevertheless, one can describe modern Composite Higgs scenarios as exhibiting one or more of the following patterns:
\begin{enumerate}
\item A Composite Higgs sector would be able to trigger Electroweak Symmetry in a non-linear fashion. This leads to new possibilities for the light Higgs boson. Its couplings, and even quantum numbers, could be different from the SM expectation. Indeed, in the SM the Higgs is  a  $SU(2)_L$-doublet, and  the way it induces EWSB completely determines how it couples to fermions and bosons, including couplings to more than one Higgs. 
On the other hand, in CHMs these couplings would be more general, even opening the possibility of a $SU(2)_L$ singlet Higgs~\cite{singletH}. This additional freedom leads to the expectation that the  SM-like Higgs, the 125 GeV state observed by the CERN collaborations, should exhibit non-standard interactions. We will discuss this first Composite Higgs pattern in Sec.~\ref{pattern1}.
\item The second set of patterns found in CHMs refers to the origin of the Composite Higgs in the spontaneous breaking of an approximate global symmetry. This breaking could lead to {\it just}  the right amount of degrees of freedom to match one single light Higgs particle and the three missing polarisations of the $W^\pm$ and $Z$ bosons. But more generically, this breaking would lead to new scalar degrees of freedom which, after partially gauging the global symmetry, could be identified  as $SU(2)_L$ singlets, doublets, or even higher representations. These non-minimal coset embeddings would lead to additional degrees of freedom which could also develop their own potential, participate in EWSB, and typically mix with the SM-like Higgs. These additional scalars would be typically required to be heavier than the Higgs, and their effects could appear at leading-order at tree-level in the effective theory via e.g., mixing with the SM-like Higgs. The study of the Run2 limits on these extensions of the SM scalar sector can be found on Sec .~\ref{pattern2}.
\item The third set of patterns found in CHMs refers to the presence of new, fermionic degrees of freedom which would assist the SM-like Higgs in its job of triggering EWSB. These fermions, often called top-partners, should be relatively light to efficiently modify the Higgs potential. They can be searched for directly, via their production at the LHC through couplings to the strong or electroweak sector, or  indirectly, as they modify the Higgs properties. The discussion on direct and indirect searches for top-partners will be developed in Sec~\ref{pattern3}.
\item In each section we also discuss situations where more than one dominant pattern could be at play in the LHC measurements. We will explain how in this case, their effect in the Higgs couplings would go in the same direction, ruling out cancellations which could invalidate the limits from the previous sections.
\end{enumerate}
 These patterns are summarised in Fig.~\ref{fig:mindmap}. 

\subsection{CHM pattern I: Non-linear effects}\label{pattern1}

 In CHM one assumes the existence of a global symmetry ${\cal G}$, spontaneously broken to a smaller subgroup ${\cal H}$. The Higgs particle and the would-be Goldstone bosons for the $W^\pm$ and $Z$ originate from this breaking. Part of ${\cal H}$ is then weakly gauged to provide the Higgs and massive gauge bosons with the correct properties under the SM. 
 
 To go beyond the SM structure of EWSB, it is useful to follow the Callan, Coleman, Wess and Zumino prescription~\cite{Callan:1969sn}. Within this prescription, one aims to build objects with definite transformation properties under $SU(2)\times U(1)$. In CHMs the object that contains the Goldstones from the global breaking ${\cal G} \to {\cal H}$ is $\Sigma = \exp{i \phi^a X^a}/f$, where $\phi^a$ are the Goldstones, $X^a$ generators in the coset ${\cal G}/{\cal H}$ and $f$ is the scale associated with the spontaneous breaking. At low energies, below $f$, one can write an effective Lagrangian involving $\Sigma$
\begin{equation}
\label{kinetic}
\mathcal L_\mathit{kinetic}^{eff} = \frac{f^2}{4} \Tr[D_\mu \Sigma^\dagger D^\mu \Sigma],
\end{equation}
which, after making a choice for weakly gauging ${\cal H}$, determines the Higgs and vector boson interactions. Assuming the Higgs is a doublet of $SU(2)_L$, similarly as in the SM, the structure that arises from this kinetic term is 
\bea
g^2 \, f^2 \, \sin^2{\left(\frac{h}{f}\right)} \ .
\eea 
One could also allow the Higgs to be a singlet under $SU(2)_L$, but that would lead to the need to impose a  tuning in the effective Lagrangian to explain the specific relations between the $Z$ and $W$ masses~\cite{Contino10, Belenetal}. We will not follow this path in this paper, and embed the CHM global symmetry structure in a way that preserves the custodial symmetry present in the SM.

Expanding around the physical Higgs boson vacuum expectation value (VEV), one finds clear predictions of the structure of the Higgs-vector boson couplings
\begin{equation}
\kappa_V = \frac{g^{CH}_{VVh}}{g^{SM}_{VVh}}= \sqrt{1-\xi} \approx 1 - \frac{1}{2}\xi \ , 
\end{equation}
where $\xi=v^2/f^2$.

Turning now to the Higgs-fermion couplings,  we note that the predictions  are not unique. In particular, the  fermion mass generation  mechanism in CHMs usually relies on the concept of partial compositeness~\cite{Contino06, Contino10}, i.e. the idea that SM fermions feel EWSB via mixing with fermionic bound states. To build structures leading to fermion couplings to the Higgs, one needs to specify the embedding of the fermionic degrees of freedom in the global symmetry structure ${\cal G}$.  As an example, in the minimal composite Higgs models, 
based on $SO(5)/SO(4)$ coset group, $\Sigma$ will transform as 5-plet of $SO(5)$ which could form a $SO(5)$ invariant either with two 5-plets or pair of 5 and 10-plets.

More general CHMs lead to very different patterns of breaking and types of embeddings for the fermion content of the model. Yes, despite spanning many model building options, it was noted in Ref.~\cite{Sanz:2017tco} that the Yukawa couplings usually fall into two choices, 
\be  \kappa_F^A = \sqrt{1-\xi}  \approx 1 - \frac{1}{2}\xi ,    \label{typeI} \ee 
and 
\be  \kappa_F^B = \frac{1-2\xi}{\sqrt{1-\xi}} \approx 1 - \frac{3}{2}\xi  . \label{typeII} \ee 

For example, the models based on coset groups  $SO(5)/SO(4)$~\cite{Agashe:2004rs,Carena:2014ria}, 
$SO(6)/SO(4)\times SO(2)$~\cite{DeCurtis:2016tsm,DeCurtis:2016scv,Mrazek:2011iu},
$SU(5)/SU(4)$~\cite{Bertuzzo:2012ya}, $SO(8)/SO(7)$~\cite{Low:2015nqa,Barbieri:2015lqa} have fermions-Higgs couplings modified by $\kappa_F^A$.
On the other hand, $\kappa_F^B$-type couplings could exist in $SO(5)/SO(4)$~\cite{Contino:2006tv, Carena:2014ria, Carmona:2014iwa, Csaki:2017cep},
$SU(4)/Sp(4)$~\cite{Gripaios09}, $SU(5)/SO(5)$~\cite{Ferretti14}, $SO(6)/SO(4) \times SO(2)$~ \cite{DeCurtis:2016tsm, DeCurtis:2016scv,Mrazek:2011iu} 
groups based models. In all these models the Higgs doublet lies within the unbroken subgroup, but in some of those there could be an extra singlet or Higgs doublet. 

A Composite Higgs, with non-standard couplings to massive fermions and bosons, would also exhibit non-standard loop-level couplings to gluons and photons ($\kappa_{g,\gamma}$)~\cite{Khachatryan:2016vau,Gillioz:2012se}  

\begin{eqnarray}  \kappa_g^2 &  = & 1.06 \kappa_t^2+0.01 \kappa_b^2-0.07 \kappa_b \kappa_t \\
\kappa_\gamma^2 & =  & 1.59 \kappa_V^2 +0.07 \kappa_t^2-0.66 \kappa_V \kappa_t  \ ,  \end{eqnarray}
as well as deviations on the Higgs width ($\kappa_H$) 
\begin{equation} \kappa_H^2  \approx   0.57 \kappa_b^2 + 0.25 \kappa_V^2 + 0.09 \kappa_g^2 . \end{equation}

Note that $\kappa_t$, $\kappa_b$ do not have to be equal (see Eqs. \ref{typeI} and \ref{typeII}) depending on how fermions are embedded within the global symmetry group.

Putting all this together, we are ready to compare measurements of the Higgs properties with  expectations from a non-linear realisation of EWSB via a Composite Higgs. The experimental inputs, described in Appendix~\ref{AppA}, are Higgs signal strengths for different production and decay channels. We compare these measurements with theoretical predictions as a function of the $\kappa$ modifiers
$\kappa_f$ , $\kappa_V$, $\kappa_g$, $\kappa_\gamma$, and $ \kappa_H$. For example, for the gluon fusion $ggH$ ($H \rightarrow \gamma\gamma$) channel, it adopts the form: $\mu^{CH}=\frac{\kappa_g^2 \kappa_{\gamma}^2}{\kappa_H^2}$.  To compare the CHM predictions with Higgs signal strength measurements from CMS and ATLAS experiments, we evaluate the $\chi^2$ statistic test
\be
\label{chiCH}
\chi^2(f) = \sum_i^{Run1, Run2} \left(\frac{\mu_i(\kappa(f))^{CH} - \mu_i^{Exp}}{\Delta \mu_i}\right)^2 \ .
\ee
Here $\mu(\kappa(f))^{CH}$, $\mu^{Exp}$ and $\Delta \mu$ denote the model prediction of the signal strength, experimental measurement, and error for
the experimental measurement, respectively. The index $i$ runs over all the measurements from Run 1 and Run 2. Note that the correlations among the different experimental measurements are not considered as they are subdominant respect to the model uncertainties introduced by the choices of $\kappa_F$.

\begin{figure}[t!]%
\begin{center}
\includegraphics[scale = 0.80]{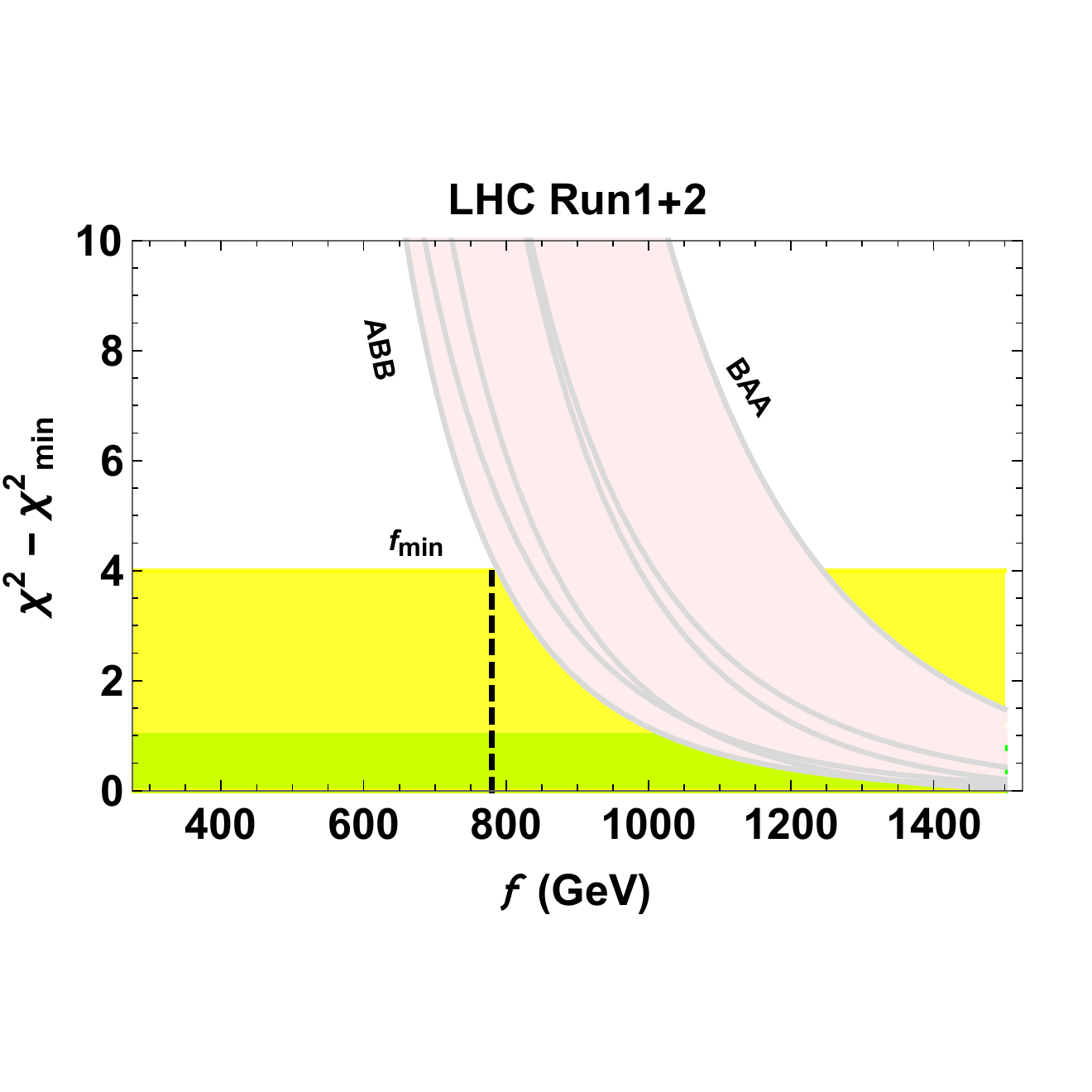}
\vspace{-2cm}
\caption{$\Delta \chi^2=\chi^2(f)-\chi^2_{min}$ for the  combination of Run 1 and 2 LHC data (signal strength measurements). The different grey lines correspond to 
different choices of fermion couplings $\kappa_F^{A,B}$ for $(\kappa_t,\kappa_b,\kappa_\tau)$. The vertical line is the lowest allowed value (95 $\%$ C.L.) of the compositeness scale ($f_{min}$) within a model where fermion couplings are of the ABB type i.e. top quark have $\kappa_A$, bottom and tau have  $\kappa_B$ factor with the Higgs couplings. Green and yellow are the 1$\sigma$ and 2$\sigma$ regions, respectively.}\label{fig:results1}
\end{center}
\end{figure} 

In this section we only focus on non-linear effects, hence the rates $\mu(\kappa(f))^{CH}$ are  sensitive to one parameter, the $\xi=v^2/f^2$ ratio, and choices for the fermion couplings from Eqs.~\ref{typeI} and \ref{typeII}. 

In Figure~\ref{fig:results1}, we show the $\chi^2$ fit of different choices for $\kappa_F$ to the combined LHC Run 1 and Run 2 data. Run 1 data is taken from the combined CMS and ATLAS analysis~\cite{Khachatryan:2016vau}. For Run 2 data, individual measurements from CMS~\cite{cmscombo,cmshtoggnew,cmshtobb,cmshtotautau,cmstthml,cmsmumu} (see table \ref{tab:cmsrun2}) and ATLAS~\cite{atlasrecent,atlasvhww} (see table \ref{tab:atlasrun2}) are considered. 
For $H \rightarrow b \bar b$ decay channel, both CMS and ATLAS have combined Run 1 and Run 2 data measurements so we have not considered the single Run 1 measurements. 

In Figure~\ref{fig:results1} we show the 1- and 2-sigma limits on $\chi^2-\chi^2_{min}$ in green and yellow, respectively. The values of $\Delta \chi^2$ depend on whether the top, bottom and tau $\kappa$ modifiers are of type $A$ or $B$. Following  Ref.~\cite{Sanz:2017tco}, we plot all the lines with all combinations and denote which combination is less constrained (ABB) and which one is constrained more strongly by LHC data, BAA. Comparing with Ref.~\cite{Sanz:2017tco}, we obtain a stronger lower bound on the scale of compositeness of  $f_{min}=$ 780 GeV. The different curves highlight the fact that there is roughly 500 GeV variation in the $f_{min}$ scale, reaching  $f_{min}\sim 1.3$ TeV for the most tightly bound scenario.

\subsection{CHM pattern II: Extended Higgs sectors}\label{pattern2}

When the group ${\cal G}$ breaks down to ${\cal H}$, light degrees of freedom are generated. Only in specific configurations one would expect exactly four Goldstones, to match the SM needs. Hence  generic  Composite Higgs scenarios would typically exhibit a pattern of Extended Higgs Sectors, with more light scalars involved in EWSB . In the simplest of such extensions, one would consider an additional singlet, see e.g. \cite{singlet}, which could mix with the SM-like Higgs doublet. 
The next simplest iteration would introduce new doublets, as is the case in e.g. the  coset group $SO(6)/SO(4)\times SO(2)$~\cite{Mrazek:2011iu}, where the light degrees of freedom organise as a Two-Higgs Doublet Model (2HDMs)~\cite{2hdm}. 

Doublets and singlets are not the only model-building possibilities which Composite Higgs models offer, e.g. on the $SU(6)/SO(6)$ coset  one has two Higgs doublets but also a custodial bi-triplet~\cite{DMCHM}. Generically speaking, a Composite Higgs scenario leading to one single doublet at low-energies is minimal but not typical, and one should consider the effect of more scalars involved in EWSB.

Moreover, this pattern opens new, interesting possibilities for model building beyond EWSB. These extra singlet or doublet pseudo-Goldstones could play a role in Inflation~\cite{DjunaCHM} or act as Dark Matter candidates~\cite{DMCHM}. In Composite Higgs scenarios, these new scalars   could  help on enhancing the strength of the electroweak phase transition~\cite{EWPTCH} by introducing new sources of CP violation and more interesting phase diagram structures, and even play some role in the QCD CP problem~\cite{Gupta:2020vxb}.  

In this section, we explore what the Run2 LHC data can tell us about these extensions. We will discuss the modification of the gauge boson and fermions couplings to the SM-like Higgs boson when the additional scalars mix with the SM-like Higgs. Their effect could also be felt at loop-level, even in the absence of a mixing, but these contributions would be suppressed by loop factors respect to mixing. Both possibilities, mixing and  loop effects, were computed and explored in Ref.~\cite{Gorbahn:2015gxa} and here we make use of these theoretical calculations and update the limits  in the context of CHMs and the possible interplay between mixing and non-linearities.

\subsubsection{Singlet scalar}

The presence of an extra singlet modifies the $\kappa_{F/V}$ by a factor of $\cos\theta_S$, where $\theta_S$ denotes the mixing angle between the neutral scalar $h$ and the extra singlet scalar field. This effect is simply due to linear mixing terms when both the singlet and the SM-like Higgs get their VEVs. Non-linearities due to the origin of the SM-like Higgs as a Composite Higgs would still be present, in exactly the same way we discussed in Sec.~\ref{pattern1}~\footnote{One could consider situations where the additional singlets or doublets do participate directly in the mechanism for EWSB, as shown in the see-saw Composite Higgs~\cite{Sanz:2015sua}, where this assumption would fail.}. Therefore, the Higgs couplings to vector bosons would be doubly modified as
\begin{equation}
\kappa_V = \cos\theta_S\sqrt{1- \xi} \approx 1 - \frac{1}{2} \xi - \frac{1}{2}\theta_S^2 \ ,
\end{equation}
where we have expanded for small values of non-linearities and mixing to show how these two effects work in the same direction, i.e. to reduce the coupling value from the SM expectation.

The non-linear part of the modification of the fermion couplings depends on the fermion embedding in representations of the global symmetry. As discussed before, we find two main choices for $\kappa_F$, namely
\begin{equation}
\kappa_F^A = \cos\theta_S\sqrt{1- \xi} \approx 1 - \frac{1}{2}\xi - \frac{1}{2}\theta_S^2 
\end{equation} 
or
\begin{equation}
\kappa_F^B = \cos\theta_S\frac{1-2\xi}{\sqrt{1-\xi}} \approx 1 - \frac{3}{2}\xi - \frac{1}{2}\theta_S^2 \ ,
\end{equation}
where again we expand for small modifications $\xi$ and $\theta_S$ to show explicitly the cooperative effort of both effects to lower the coupling.

\begin{figure}[t!]%
    \begin{center}
\includegraphics[scale = 0.80]{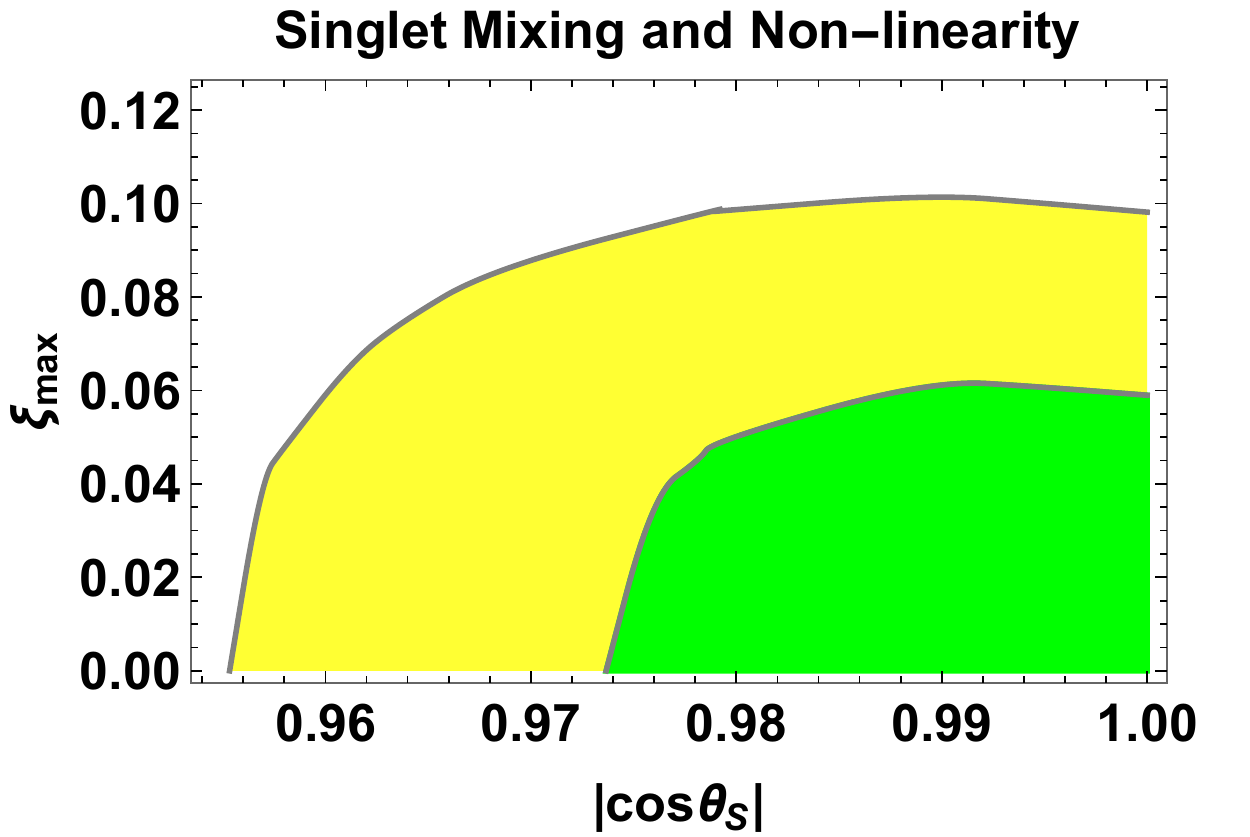}
    \end{center}
    \caption{The bound on the $\xi_{max}$ as a function of $\cos\theta_S$ in case of singlet mixing scenario for the ABB type model.}\label{fig:results2}
\end{figure}

We  perform the fit to the Higgs data as in Sec.~\ref{pattern1} but now including the effect of mixing, $\chi^2(f,\theta_S)$. As shown in Figure~\ref{fig:results1}, we vary options for the fermion couplings and find the $f_{min}$. We perform this fit for different values of the mixing angle, always finding the minimum $f$ corresponding to a set of options for $\kappa_F$. 
The result of this procedure is shown in Figure~\ref{fig:results2}, where we plot $\xi_{max} = v^2/f^2_{min}$ scale  as a function of the singlet mixing. On the right side of the plot we see the effect of small mixing, where we recover the result from the previous section. As we move towards the left in the plot, the mixing becomes more important and at some point $\xi_{max}\to 0$, we obtain the pure mixing limit of $|\cos{\theta_S}|\simeq$0.96. 

Note that the mixing angle is determined by singlet mass ($m_S$),  singlet vev ($v_S$) and  quadratic coupling of light Higgs doublet and singlet scalar ($\lambda$), as given below 
\be \cos\theta_S=\cos\left( \frac{v^2}{(m_S^{eff})^2}\right) \quad ; \quad m_S^{eff}=\frac{m_S}{\sqrt{\lambda (v_S/v)}}\ee
where $v=246$ GeV.
Therefore, the limit $|\cos{\theta_S}|\simeq$0.96 would be equivalent to $m^{eff}_S \gtrsim$ 450 GeV.

In between these two asymptotic limits, the  constraints  on the scales $f$ and $m^{eff}_S$ become stronger, as both mixing and non-linearities work together to reduce the Higgs couplings.

\subsubsection{Two Higgs Doublet Models (2HDM)}
\begin{table}[t!]
\begin{center}
\begin{tabular}{|c|c|c|}
\hline
Model & $\kappa_V$ & $\kappa_F$ \\
\hline
$\mathrm{Type}-\mathrm{I}$ & $s_{\beta-\alpha}$ & $\kappa_u = \kappa_d = \kappa_{\ell} = c_{\beta-\alpha} / t_\beta+s_{\beta-\alpha}$  \\
$\mathrm{Type-II}$ & $ s_{\beta-\alpha}$ &  $\kappa_u = c_{\beta-\alpha} / t_\beta+s_{\beta-\alpha}$ \\
 & & $\kappa_d = \kappa_{\ell} =s_{\beta-\alpha} - t_\beta c_{\beta-\alpha}$  \\
$\ell-\mathrm{Specific}$ & $s_{\beta-\alpha}$ & $\kappa_u = \kappa_d = c_{\beta-\alpha} / t_\beta+s_{\beta-\alpha}$ \\ 
& & $\kappa_{\ell}= s_{\beta-\alpha} - t_\beta c_{\beta-\alpha}$ \\
$\mathrm{Flipped} $ &  $s_{\beta-\alpha}$ & $\kappa_u = \kappa_{\ell} = c_{\beta-\alpha} / t_\beta+s_{\beta-\alpha}$ \\
& &  $ \kappa_{d} = s_{\beta-\alpha} - t_\beta c_{\beta-\alpha}$  \\
\hline
\end{tabular}
\end{center}
\caption{The $\kappa_V$ and $\kappa_F$ expressions in Two Higgs doublet models. Here $s_x/c_x/t_x=\sin x/\cos x/ \tan x$.}
\label{kappa2hdm}
\end{table}

As discussed in the previous section, the mixing effect due the additional Higgses in a Composite Higgs model does not typically couple with the non-linear effects at leading order. 
Hence, in composite two Higgs doublet models the vector and femion couplings to the Higgs, $\kappa_{V/f}$, would still adopt this factorisable form $\kappa^{2HDM}_{V/F} \times \kappa^{CH}_{V/F}$, where $\kappa_{V/F}^{CH} $ has been discussed in Sec.~\ref{pattern1} and 
$\kappa^{2HDM}_{V/F}$ correspond to the modifications of the SM-like Higgs couplings in 2HDM models. 

The explicit form of $\kappa^{2HDM}_{V/F}$ is given in Table~\ref{kappa2hdm} for various types of 2HDM models. In these scenarios, vector and fermion modifiers are determined by two parameters, the ratio of symmetry breaking VEVs of Higgs 
doublets ($\tan\beta$) and neutral Higgs mixing angle ($\alpha$).

\begin{figure*}[t!]%
    \begin{center}
 \includegraphics[scale = 0.55]{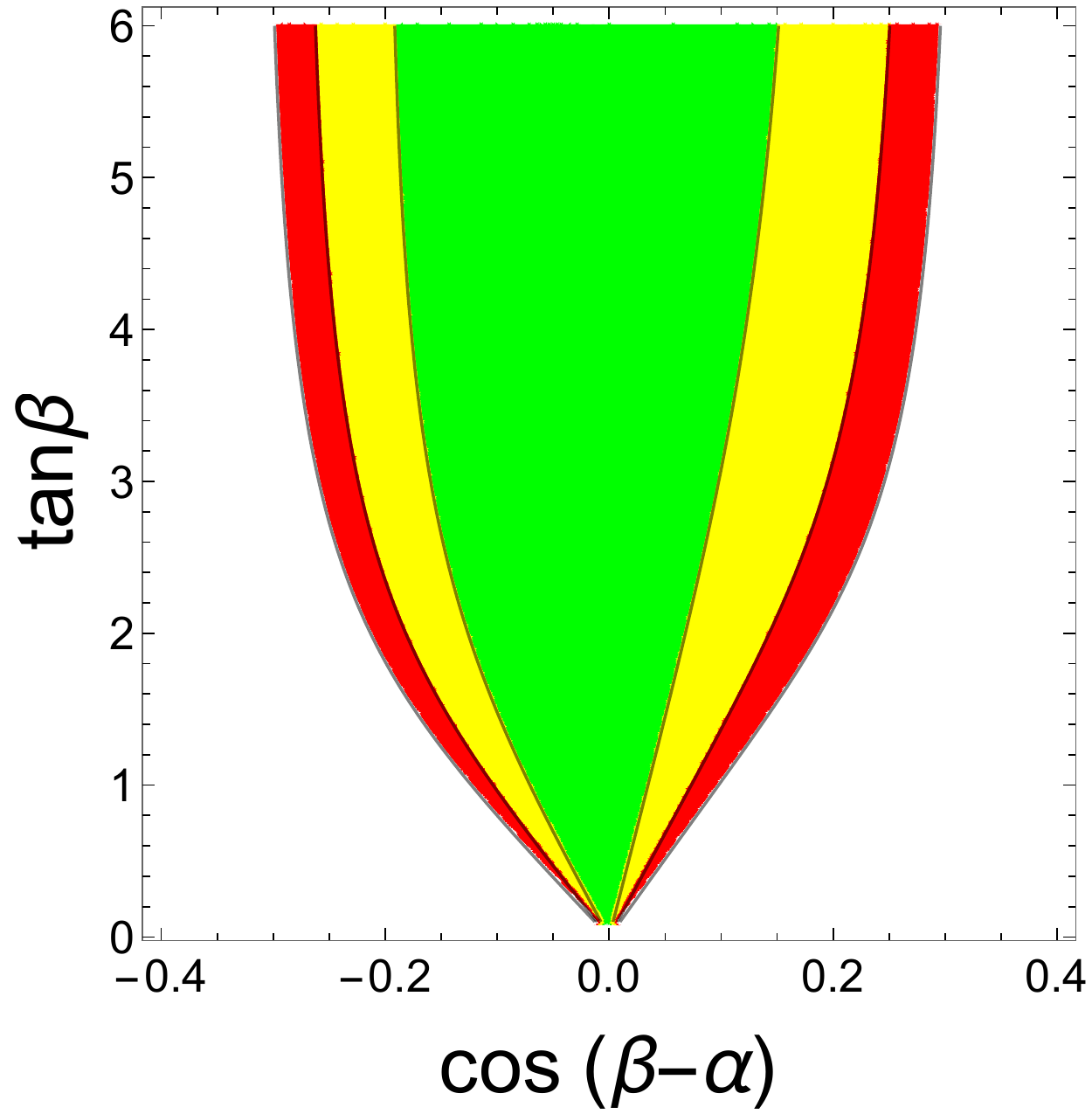}
 \includegraphics[scale = 0.55]{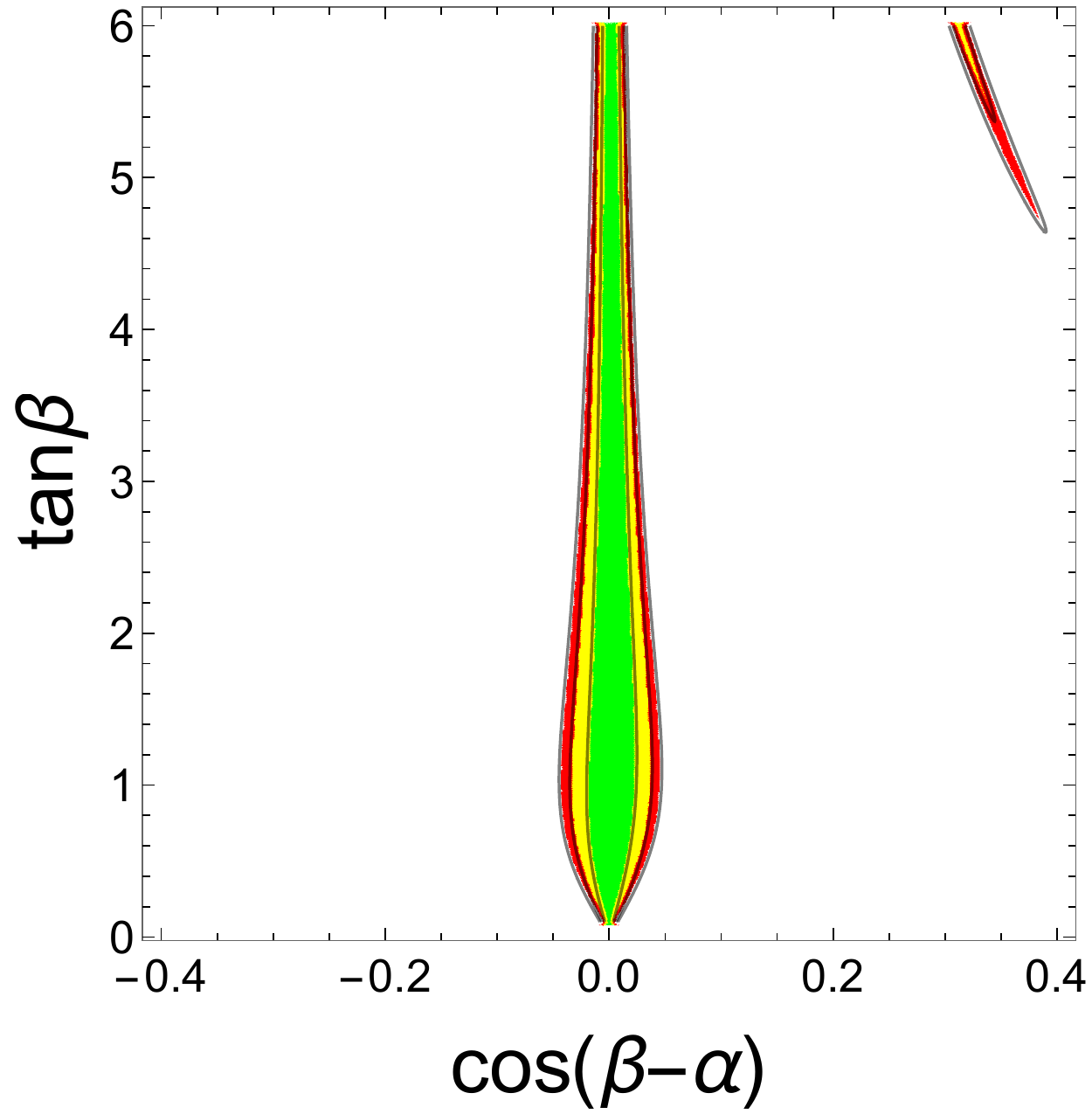}
    \end{center}
    \caption{The 68 $\%$(green), 95 $\%$(yellow) and 98 $\%$(red) C.L. limits in the 
    $\cos(\beta-\alpha)-\tan\beta$ plane for 2HDM Type I (left plot) and Type II (right plot) with $\xi=0$.}\label{fig:results3}
\end{figure*} 

In our analysis, we only consider the main scenarios, Type I and Type II. The $\chi^2_{2CHM}$ statistical test for a Composite Higgs with possible mixing with another doublet becomes a function of three parameters: $f, \cos ({\beta-\alpha}),$ and $\tan\beta$. 

\begin{figure*}[t!]%
    \begin{center}
\includegraphics[scale = 0.52]{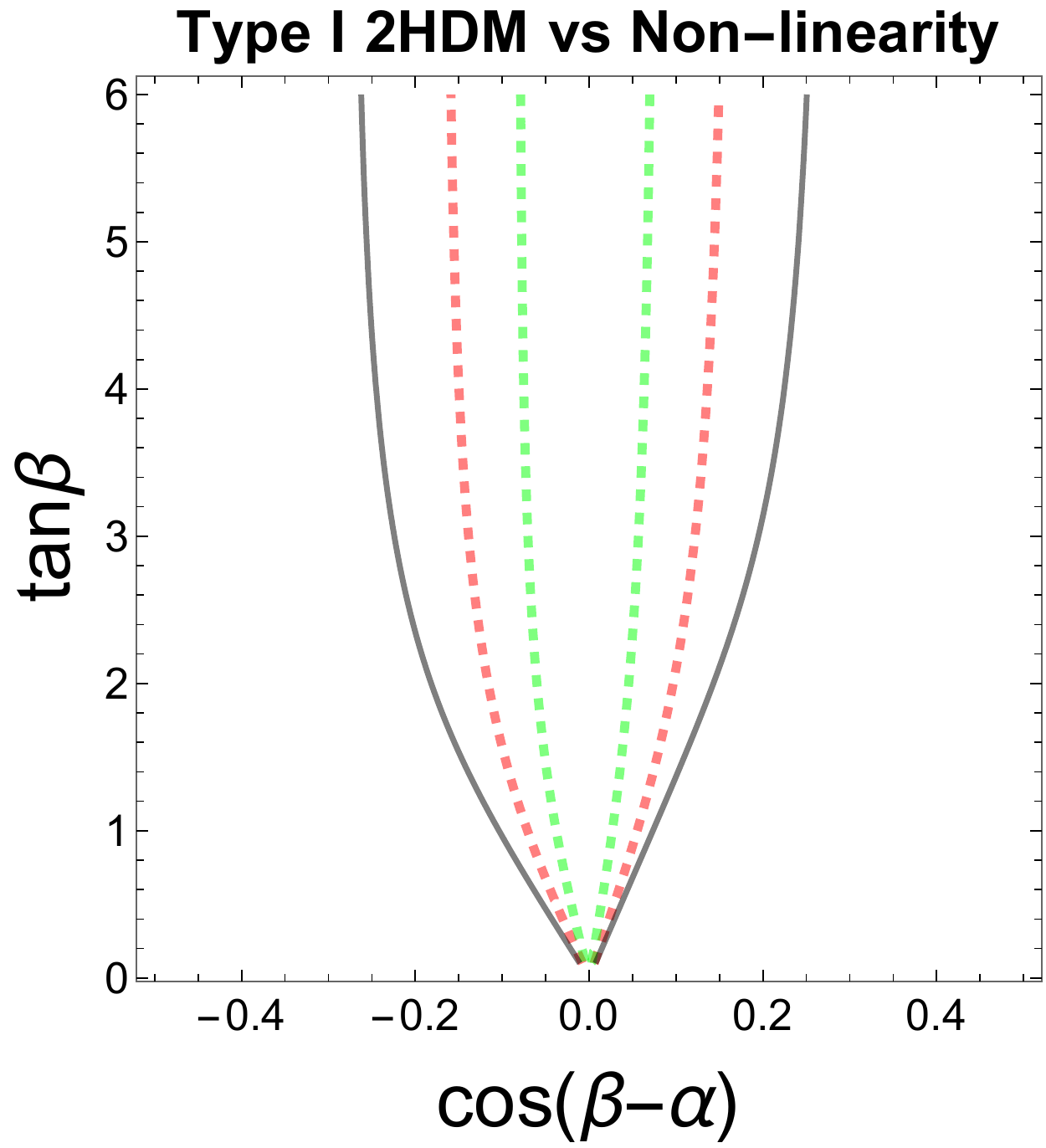}
 \includegraphics[scale = 0.55]{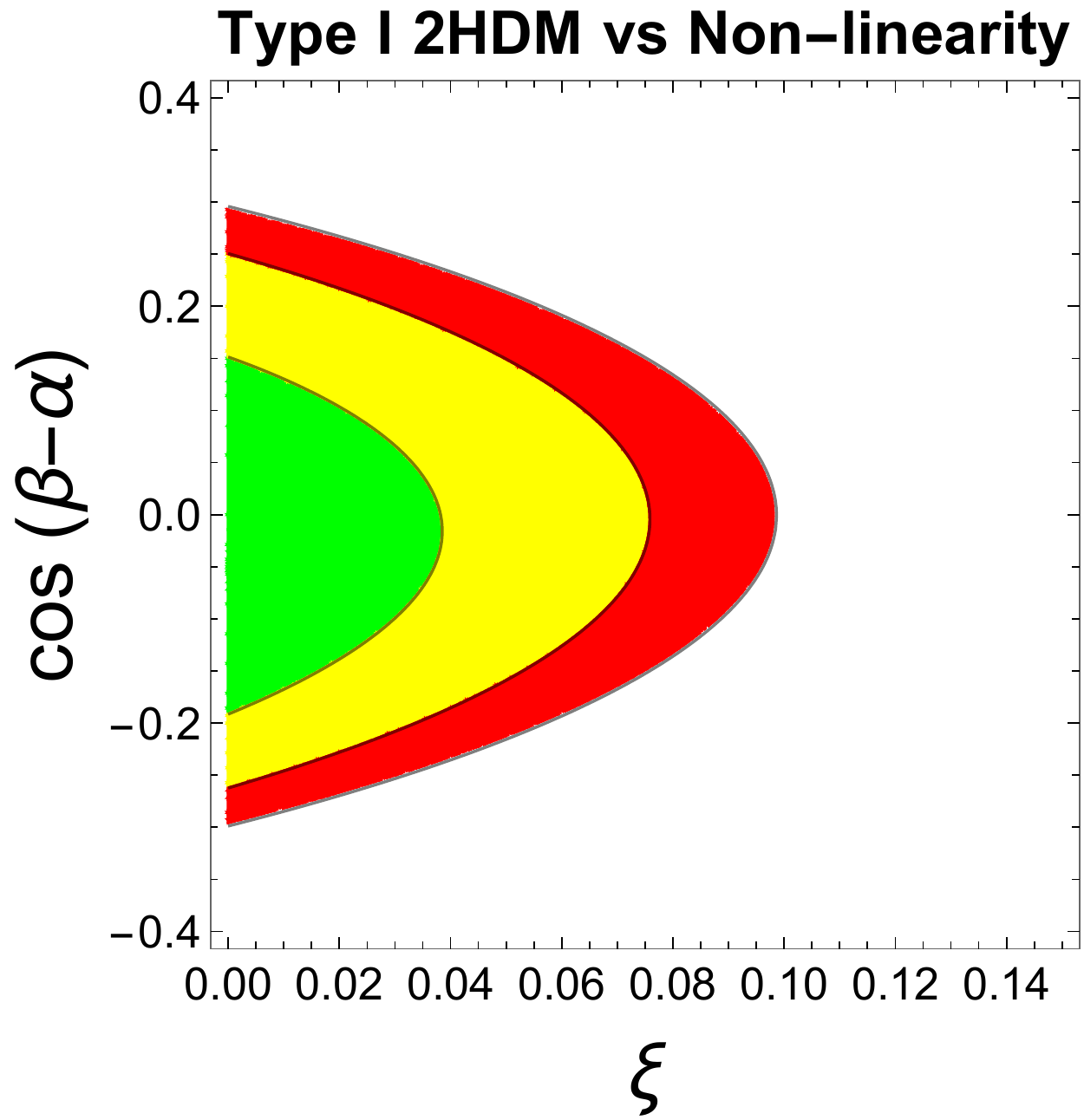}
      \end{center}
    \caption{Type I 2HDM results. {\it Left plot: }The 95 $\%$ C.L. limit including non-linear effects. The black, red and green lines correspond to choices of the parameter $\xi=0$, 0.05 and 0.07, respectively. {\it Right plot: } The 68 $\%$(green), 95 $\%$(yellow) and 98 $\%$(red) C.L. limits in the $\xi$-$c_{\beta-\alpha}$ plane for $\tan\beta=6$.}\label{fig:results4}
\end{figure*} 

First we discuss the information that Run2 LHC data provides on type I and II 2HDMs, without considering the non-linear effects. This information is shown in Figure~\ref{fig:results3}, where the coloured regions green, yellow and 
red represent 68 $\%$, 95 $\%$ and 98 $\%$ C.L. limits in the $\cos(\beta-\alpha)-\tan\beta$ plane, respectively. As expected, the SM-like Higgs prefers a region with $\cos{\beta-\alpha}\simeq$ 0, the alignment limit~\cite{haberailignment}. 

The alignment limit can be understood in terms of the potential parameters in the 2HDM, couplings between the two doublets $\tilde \lambda_i$ and masses $\tilde \mu_{1,2}$, see Ref.~\cite{Gorbahn:2015gxa} for more details. In terms of these parameters,  the deviation from the alignment limit can be parametrised as $\cos{\beta-\alpha} \sim \lambda_6 v^2/\tilde \mu_2^2$.

Next we introduce the additional non-linear effects in the left panel in Figure~\ref{fig:results4}, where we show how  the 95 $\%$ C.L. regions are modified with non-linear effects. The non-linear effects, as in the case of the singlet add up to the doublet mixing. Larger non-linearities restrict further the parameter space of 2HDMs. 

Finally, as the $\tan\beta$ dependence is rather mild for $\tan\beta \gtrsim$ 2, we fix the value of $\tan\beta=$6 to explore the interplay between $\cos{\beta-\alpha}$ and $\xi$. In the right panel of Figure~\ref{fig:results4}, we plot the one-, two- and three-$\sigma$ contours in this plane. In the alignment limit, $\cos{\beta-\alpha}\simeq$ 0, we recover the pure non-linear limit $\xi\simeq 0.08$ at 95\%C.L.. As we move from the alignment limit, and allow a larger amount of doublet mixing, the limit on $\xi$ becomes milder, leading to a stronger bound on the compositeness scale $f$. On the left side of the plot  where $\xi\to $ 0,  we find the limit on non-decoupling effects $\cos{\beta-\alpha}\simeq$ 0.25 at 95\%C.L., which would correspond to the scale for the second doublet $\tilde \mu_2 \sim $ 500 GeV for $\tilde \lambda_6 \sim {\cal O}(1)$.

\subsection{CHM pattern III: Extended fermionic sector}\label{pattern3}

We finish this section on patterns, analysing a typical building block in Composite Higgs scenarios: the presence of new fermionic degrees of freedom, not too far from the electroweak scale. These fermionic degrees of freedom are usually coming along as top partners, new vector-like fermions which mix with the SM top quark and modify its couplings with the Higgs.

We first focus on the mixing, and present results for the $SU(2)_L$-singlet top partner scenario $T^0_{L,R}$. We calculate the indirect bound on these states from measurements on the differential distributions of the the boosted Higgs in association with an energetic jet. The methodology used here was developed in Ref.~\cite{BanfiSanzHandj} and further expanded in Refs.~\cite{andreaHj} and \cite{BanfiBondSanzMartin}. We will then evaluate other sources of experimental constraints, including direct searches for the new state.

To describe the effect of the top-partner in the Higgs behaviour we need to specify the mixing parameters between the top-partner and the SM top quark. For the singlet case, the mass matrix of the SM top quark and 
Dirac fermion top partner $T=(T_L,T_R)$ can be written as
\be \begin{pmatrix} \bar t_L & \bar T_L \end{pmatrix}  \begin{pmatrix} \frac{y_t h}{\sqrt{2}} & \Delta \cr 0 & M \end{pmatrix} \begin{pmatrix}  t_R \cr T_R  \end{pmatrix} \ ,\ee
where $\Delta$ describes the mixing between $t$ and $T$, and $M$ is the Dirac mass of the top partner. After diagonalising this matrix by a bi-unitary transformation, 
we can calculate the mass eigenstates  $M_T$ and $m_t$  and the mixing angle 
\be \theta_R=\frac{1}{2} \sin^{-1} \left( \frac{2 m_t M_T \Delta}{(M_T^2-m_t^2)M} \right)   \ . \ee
Note that $ \theta_R$ and $ \theta_L$ are related in a simple way,
 \be \tan \theta_L= \frac{M_T}{m_t}  \tan \theta_R  .\ee

\subsubsection{Top-partner indirect searches}~\label{topindirect}

Contrary to the previous sections, total Higgs rates and their respective limits on vector and fermion couplings $\kappa_{V/F}$ would not be sensitive to the presence of the top partner. On the other hand, high-$p_T$ probes would access the top partner mixing and mass scale indirectly~\cite{BanfiSanzHandj,andreaHj}.  

To analyse the effect of $T$ in the Higgs production in association with radiation, it is useful to define a quantity which depends on $M_T$ and $\theta_R$ based on semi-differential measurements,
\be  \sigma (p_T > p_T^{cut})=\int_{p_T^{cut}}^{\infty} dp_T \frac{d \sigma}{d p_T} \ . \ee 

In particular, we define a quantity based on these differential measurements,
\be  \delta (p_T^{cut},M_t,\sin\theta_R)= \frac{\sigma_{t+T} (p_T^{cut},M_T,\sin\theta_R)-\sigma_t ( p_T^{cut})}{\sigma_t (p_T^{cut})} \ , \ee
which exhibits good properties from the point of view of systematic and statistical fluctuations, see Ref.~\cite{BanfiSanzHandj} for more details.

\begin{figure*}[t!]%
    \begin{center}
  \includegraphics[scale = 0.67]{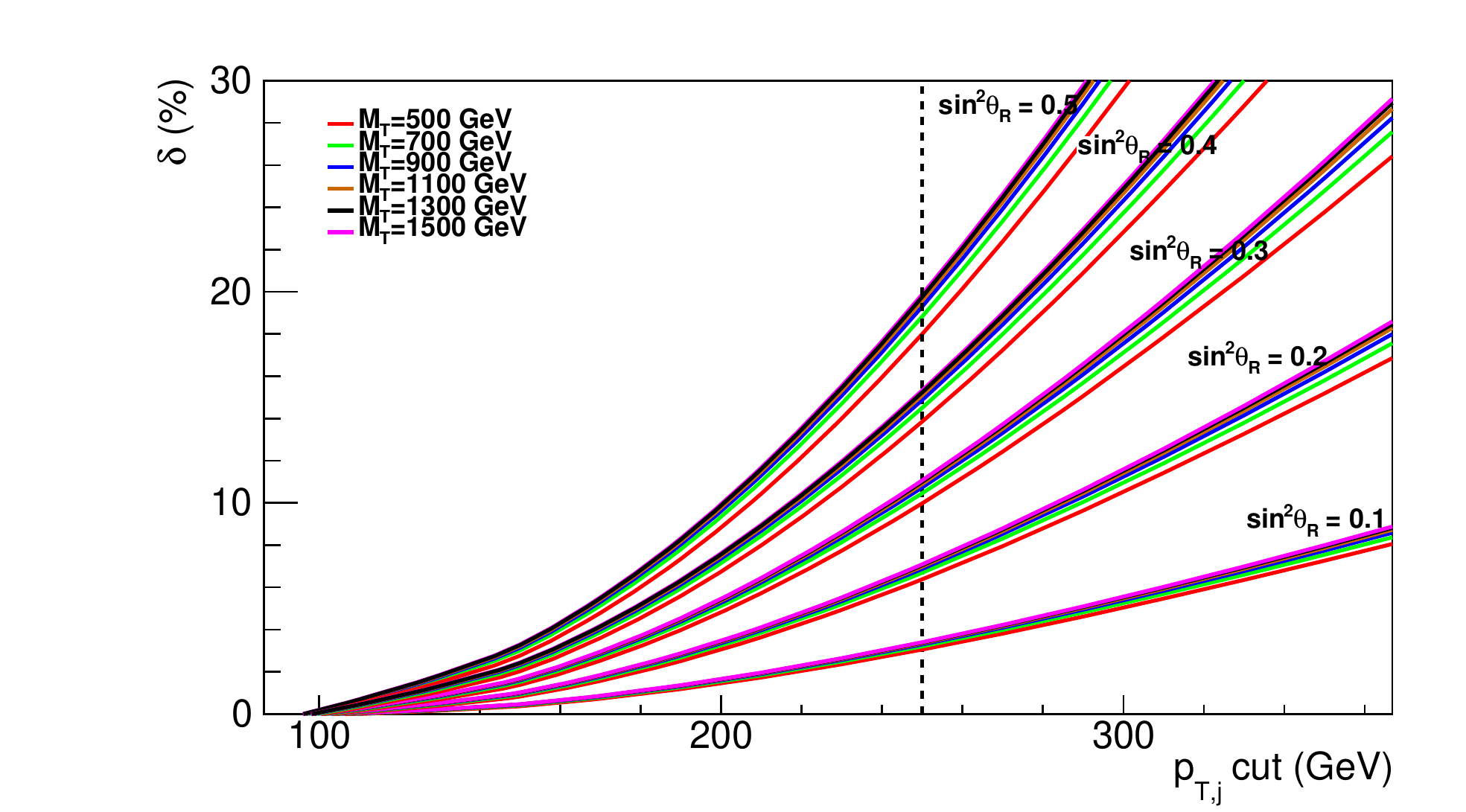}
\end{center}
\caption{Percentage enhancement in the Higgs$+$ jet cross-section by the top partner contribution as a function of $p_T^{cut}$ for different values of mixing angle and $M_T$.}\label{fig:DeltaVsPt}
\end{figure*} 

With this observable as an indirect probe for fermionic top-partners, we calculated $\sigma_t$ and $\sigma_{t+T}$ for $\sqrt{s}=$13 TeV by varying $M_T$ and $\sin \theta_R^2$. The results can be seen in Figure~\ref{fig:DeltaVsPt}, where we plot the relative change in the Higgs+jet cross-section as a function of $p_T^{cut}$ for different combinations of the mixing angle and $M_T$. For this plot, $M_T$ is varied from 500 GeV to 1500 GeV, and one can see that $\delta$ has a weak dependence on $M_T$. We can then use the experimental measurements in this channel to put a bound on the mixing angle. In particular, we used the ATLAS Higgs differential cross-section analysis~\cite{ATLASdiphotondiff}, which reports an 8$\%$ deviation at 95 $\%$ C.L. in the 250-350 GeV (diphoton) $p_T$ bin from the SM cross-section. We can then look back at our Fig.~\ref{fig:DeltaVsPt} and note the variation of $\delta$ as a  function of $\sin\theta_R$  for  $p_T^{cut}$=250 GeV. 

\begin{figure}[t!]%
    \begin{center}
\includegraphics[scale = 0.80]{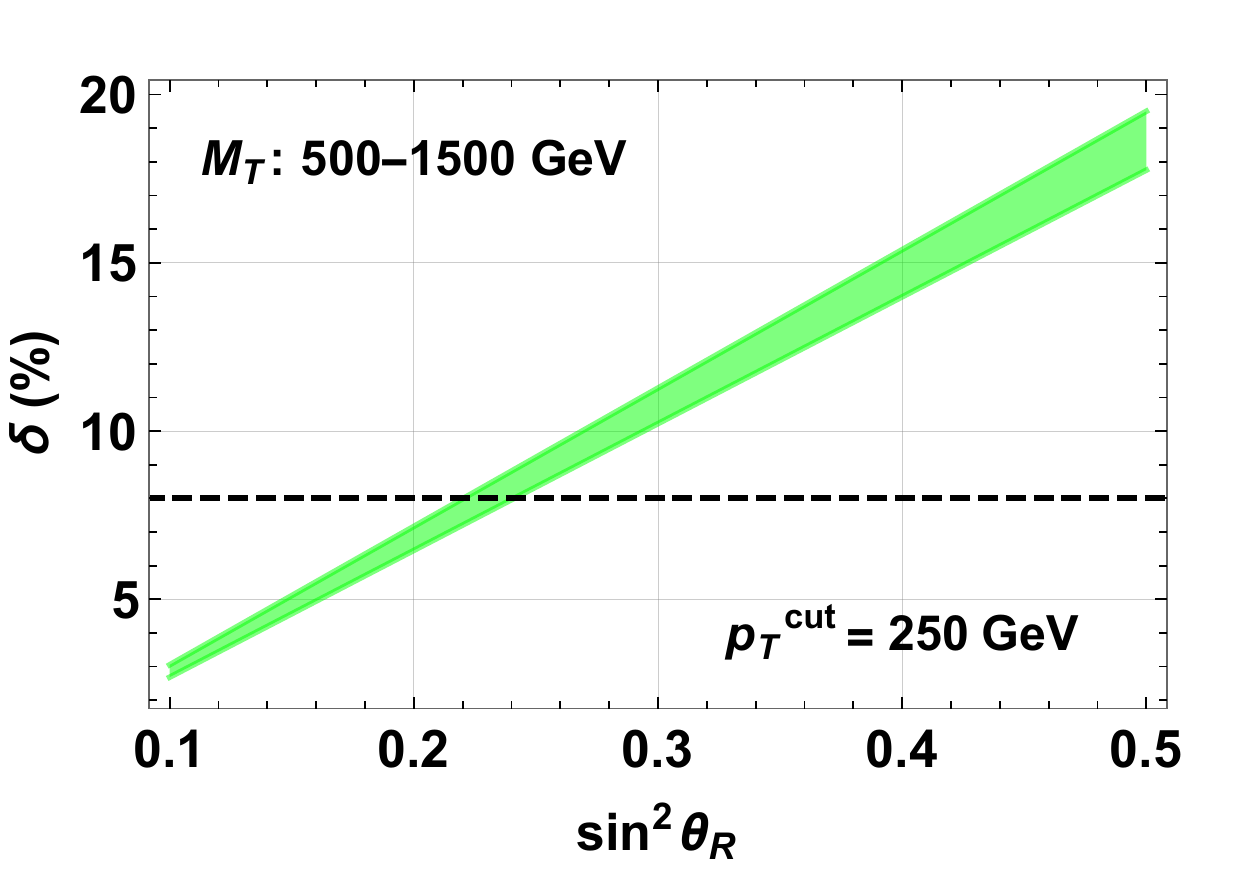}
\end{center}
\caption{Percentage enhancement in the Higgs+jet cross-section by the top partner contribution ($\delta$) as a function of the top-partner mixing angle for $p_T^{cut}$=250 GeV. The horizontal hashed line corresponds to the experimental bound~\cite{ATLASdiphotondiff}, and the band width to the variation in $M_T\in $[500,1500] GeV.}\label{fig:thetaRvsDelta}
\end{figure}

In Fig.~\ref{fig:thetaRvsDelta} we show the variation of $\delta$ as a  function of $\sin\theta_R$ for a fixed $p_T^{cut}=250$ GeV. The  green band corresponds to varying $M_T$  from 500 GeV to 1500 GeV. The 8\% limit is noted with a hashed black line, which allows us to set a limit on $\sin^2 \theta_R \lesssim 0.25$.

Another source of constraints from a singlet VLQ top-partner would come from a SM Effective Field Theory (SMEFT) global analysis of Run2 LHC and LEP measurements. The singlet top-partner would produce a SMEFT pattern characterised by relations among some SMEFT operators, 
whereas all the other operators would be zero at tree-level, see Ref.~\cite{GranadaSMEFT} for a dictionary between many extensions of the SM and their SMEFT matching. See also the discussion in Ref.~\cite{CHeftrecent} on the top-partners and the SMEFT framework.
Using this dictionary and a global SMEFT analysis, in Ref.~\cite{Ellis:2020unq} the parameters of the singlet top-partner were bound to $\sin{\theta_L}^2 \lesssim 0.05$ for $M_T \simeq $ 1 TeV. 

\subsubsection{Top-partner direct searches}
Boosted Higgs measurements provide one handle on top-partners.  The LHC New Physics search programme includes very mature analyses looking for single and pair production of  Vector-Like Quarks (VLQ).  Pair production of VLQs is a dominantly strong interaction process, whereas single production relies on electroweak couplings. 

A singlet up-type VLQ (T) is excluded up to 1.31 TeV
by the ATLAS $\sqrt{s}=13$ TeV (36.1 fb$^{-1}$) search for pair produced VLQs~\cite{AtlaspairprodVLQs} after considering all decay modes. This limit is a combined limit from several searches for the pair production of VLQs. 

From the CMS side, there are separate limits from each  analysis of pair production of VLQs. The exclusion limit on singlet T from dilepton final state of VLQs decaying to Z boson is 1280 GeV~\cite{CMSVLQpairDilepton}. The CMS analysis for all leptonic final states excludes $M_T$ in the range 1140-1300 GeV corresponding to different branching fractions~\cite{CMSVLQpairAllleptonfinalstates}, and the analysis of the  $bW\bar{b}W$ final state excludes singlet $T$ up to 1295  GeV~\cite{CMSVLQpairbwbw}. The analysis targeting fully hadronic final states provides a bound up to 1.3 TeV for a specific combination of the branching fractions of VLQs~\cite{CMSVLQpairfullthad}.
 
\begin{figure}[t!]%
    \begin{center}
  \includegraphics[scale = 0.65]{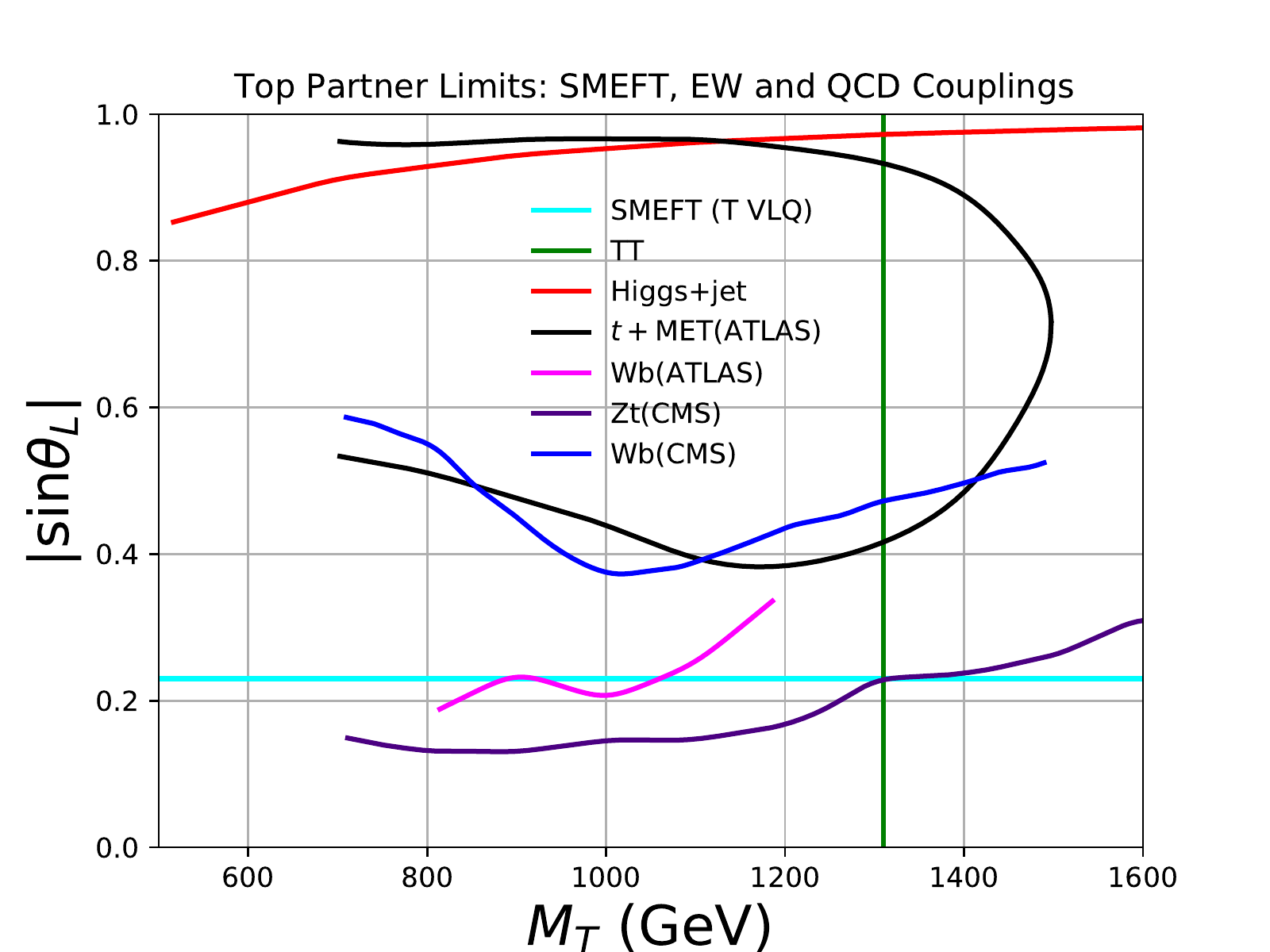}
\end{center}
\caption{Direct searches bounds on singlet top-partners from single and double production of VLQs, and indirect bounds from the Higgs+jet final state. Blue, orange, green and violet curves correspond to the ATLAS single VLQ bound from $Wb$ channel~\cite{singleVLQsatlaswb}, ATLAS bound from top+MET final state~\cite{AtlastopMET}, CMS $Wb$ final state~\cite{CMSsingleVLQbw} and CMS $Zt$ final state~\cite{CMSsingleVLQZt}, respectively. The vertical line corresponds to the ATLAS pair production bound~\cite{AtlaspairprodVLQs}. The SMEFT bound is taken from Ref.~\cite{Ellis:2020unq} and the red line indicates the limit from Higgs+jet differential distributions discussed in Sec.~\ref{topindirect}.}\label{fig:toppsummary}
\end{figure} 

The single VLQs production process is mediated by its coupling with SM particles, hence it provides a bound on the mixing as a function of the mass of VLQs~\cite{singleVLQsatlaswb,AtlastopMET,CMSsingleVLQbw}. Both the production cross-section and decay branching ratios are sensitive to mixing effects. The communication of the single-production results is a bit more cumbersome than for double-production. In some analyses, this mixing is directly parametrised by the mixing angle ($\sin\theta_{L,R}$) and sometimes by the coupling $c^{bw}$ where $c^{bw}_{L/R}=\sqrt{2} |\sin\theta_{L/R}|$~\cite{HandbookVLQs} is just a re-scaling of the mixing angle. ATLAS analysis cross-section limits are interpreted in terms of either $|\sin\theta_L|$, or both $C_W^2 =(c^{bw}_L)^2+(c^{bw}_R)^2$ and $|\sin\theta_L|$. On the other hand, CMS exclusion limits on production cross-section times branching fraction are  provided only as a function of  $m_{VLQ}$.  Additionally, the theoretical cross-section is  provided for the fixed value of $BR$ and $c^{bw}=0.5$. We use this additional information to calculate the approximate bound on $c^{bw}$ as a function of $m_{VLQ}$. 

We translate CMS bound in terms of mixing angle $|\sin\theta_L|$ by parametrizing the production cross-section as
 $\sigma (T) \times BR(T\rightarrow bW) = \tilde \sigma(T) {c^{bw}}^2 \times BR(T\rightarrow bW)$, where $\tilde\sigma(T)$ is calculated from the theoretical cross-section with fixed $c^{bw}$. Then using the bound on cross-section (including BR), we calculate the limit on $c^{bw}$ which we further translate to $|\sin\theta_L|$ using the relation $c^{bw}_{L/R}=\sqrt{2} |\sin\theta_{L/R}|$. 
 Note that we have not considered the effect of $C_W$ variation on the width of $T$. We also consider other decay channels of the singlet vector-like top to $ht$ and $Zt$. Note that CMS analyses for $Wb$~\cite{CMSsingleVLQbw} and $Zt$~\cite{CMSsingleVLQZt} final states do assume 100 $\%$ branching ratio, and the ATLAS collaboration has considered $Wb$  and top +MET final states~\cite{AtlastopMET} assuming BR($T\rightarrow Zt)=0.25$ from the singlet model. 
 
In Figure~\ref{fig:toppsummary} we show all these direct and indirect limits together in the plane of top-partner mass versus mixing angle.  Double production of VLQs (dark-green vertical line) sets a limit on the mass $M_T \gtrsim$ 1.3 TeV. The SMEFT limit (light blue horizontal line) is very strong, and independent of the top-partner mass as long as $M_T \geqslant v$. The current sensitivity of the other indirect probe for top-partners, the $H$+jet channel, is below the SMEFT fit, although it is worth noticing that our $H$+jet analysis is not at the same level of sophistication as the SMEFT's~\cite{Ellis:2020unq}. Note that for the ATLAS top+MET analysis (black line) the $|\sin\theta_L|$ values covered by the closed contour are excluded.


\section{Conclusions\label{conclusions}}

In this work we have studied the impact of the LHC Run 2 measurements on general Composite Higgs scenarios. 

The dataset we used includes Higgs signal strength measurements from CMS and ATLAS, differential properties of boosted Higgs production, several searches for single- and double-production of vector-like quarks, and a specific result from a global fit to SMEFT properties obtained in Ref.~\cite{Ellis:2020unq}.

We have described the most significant deviations expected from Composite Higgs scenarios, going from the simplest effects, non-linear couplings, to the most complex interplay between direct and indirect searches for additional fermionic  degrees of freedom.  

We have shown how in general Composite Higgs scenarios one should expect  simultaneous effects from more than one source. Typically, the composite nature of the Higgs is tied to the non-linear realisation of EWSB and the  scale of compositeness $f$. But generic Composite Higgs scenarios exhibit a richer phenomenology, in particular new light scalars or fermions are also typical predictions of these scenarios. 

These new degrees of freedom, if close enough to the electroweak scale, will also modify the Higgs couplings. For example, mixing of additional scalars participating in EWSB will add to non-linearities to reduce even further the couplings of the Higgs to massive particles, leading then to even stronger limits on the scale of compositeness $f$. Note that  this scale is tied to the mass of new vector resonances like $W'$ and $Z'$, and to the degree of tuning of Higgs potential. The Run2 LHC data has pushed this scale into the TeV range, and we have shown that more complex scenarios would push this tuning even further. Hence Run3 and future LHC runs should  have a good handle at natural Composite Higgs scenarios.  

\acknowledgments
The work of V.S. by the Science Technology and Facilities Council (STFC) under grant number ST/P000819/1. We want to thank Stephan Huber and Jack Setford for many fruitful discussions during the early stages of this work. We thank Andrea Banfi for the help regarding running the Higgs+jet process in his code implementation.
\appendix

\section{Data inputs}~\label{AppA}
\begin{table}
\begin{center}
 \begin{tabular}{|c| c |c | c |}
 \hline
Production & Decay Channel & Lumi[Reference] &  Signal Strength \\
\hline 
 \hline
  ggH & $H \to ZZ$ & {137.1 fb$^{-1}$}\cite{cmscombo}&     $0.98^{+0.12}_{-0.11}$ \\
 VBF &  & &  $0.57^{+0.46}_{-0.36}$  \\
VH & &   &  $1.10^{+0.96}_{-0.74}$\\
 $t \bar t H, tH$ & &  &  $0.25^{+1.03}_{-0.25}$ \\
\hline 
ggH & $H \to \gamma \gamma $ & {137 fb$^{-1}$}\cite{cmshtoggnew}  &  $0.98^{+0.13}_{-0.10}$ \\
 VBF & & & $1.15^{+0.36}_{-0.31}$ \\
 VH &  & &   $0.71^{+0.31}_{-0.28}$  \\
 $t \bar t H$ & & &  $1.40^{+0.33}_{-0.27}$ \\
\hline
 
ggF & $H \to WW $ & {35.9 fb$^{-1}$}\cite{cmscombo} & $1.28^{+0.20}_{-0.19}$  \\
VBF & & & $0.63^{+0.65}_{-0.61}$  \\
WH & & & $2.85^{+2.11}_{-1.87}$  \\
ZH & &  & $0.90^{+1.77}_{-1.43}$ \\
ttH & &  & $0.93^{+0.48}_{-0.45}$ \\
\hline
ggH & $H \to b \bar b $ & upto 77.4 fb$^{-1}$\cite{cmscombo}& $2.45 ^{+2.53}_{-2.35}$ \\
WH &  &  &   $1.27^{+0.42}_{-0.40}$  \\
ZH &  &  &   $0.93^{+0.33}_{-0.31}$  \\
$t \bar t H$ & & &  $1.13^{+0.33}_{-0.30}$ \\
\hline
 VBF & $H \to  b \bar b $ & upto 77.2fb$^{-1}$\cite{cmshtobb} & $2.53 \pm {1.53}$ \\ \hline
ggH & $H \to \tau \bar \tau $ & {137 fb$^{-1}$}\cite{cmshtotautau} &  $0.98^{+0.20}_{-0.19}$ \\
VBF+V(qq)H & &  & $40.67^{+0.23}_{-0.22}$ \\
\hline
 $t \bar t H$ & $H \to ML$  & { 137 fb$^{-1}$} \cite{cmstthml} &  $0.92^{+0.26}_{-0.23}$\\
\hline
ggH &  $H \to \mu\mu$  & {137 fb$^{-1}$} \cite{cmsmumu} & $0.63^{+0.65}_{-0.64}$ \\
VBF &  & & $1.36^{+0.69}_{-0.61}$ \\ 
VH &  & & $5.48^{+3.10}_{-2.83}$ \\ 
$t \bar tH$  & & & $2.32^{+2.27}_{-1.95}$ \\ \hline
\end{tabular} 
 \end{center}
  \caption{Summary of CMS Run 2 Higgs signal strength measurements.}
  \label{tab:cmsrun2}
\end{table}

\begin{table}
\begin{center}
 \begin{tabular}{|c|c|c|c|}
 \hline
Production & Decay Channel & Lumi[Reference]& Signal Strength \\
\hline
ggF & $ H \to ZZ $ &{139 fb$^{-1}$}\cite{atlasrecent} &  $0.94^{+0.11}_{-0.10}$  \\
VBF&   & &  $1.25 ^{+0.50}_{-0.41}$ \\
VH&     &  &  $1.53^{+1.13}_{-0.92}$  \\
\hline
ggF  & $H \to \gamma \gamma $ & {139 fb$^{-1}$} \cite{atlasrecent}&  $1.03\pm {0.11}$ \\ 
 VBF &  & &  $1.31^{+0.26}_{-0.23}$ \\
VH &   & &   $1.32^{+0.33}_{-0.30}$  \\
$ttH+tH$ &   & &  $0.90^{+0.27}_{-0.24}$ \\
\hline
ggF & $ H \to WW $ &{36.1 fb$^{-1}$}\cite{atlasrecent} &  $1.08^{+0.19}_{-0.18}$ \\
VBF&  &{36.1 fb$^{-1}$} \cite{atlasrecent}&  $0.60^{+0.36}_{-0.34}$  \\
VH&  & {36.1 fb$^{-1}$}\cite{atlasvhww}&  $2.5^{+0.9}_{-0.8}$ \\
\hline 
VBF & $ H \to b  b $ & {24.5-30.6 fb$^{-1}$}\cite{atlasrecent}&  $3.03^{+1.67}_{-1.62}$ \\
VH&    & {139 fb$^{-1}$}\cite{atlasrecent} &  $1.02^{+0.18}_{-0.17}$  \\
$ttH+tH$&   & {36.1 fb$^{-1}$} \cite{atlasrecent}&  $0.79^{+0.60}_{-0.59}$  \\
\hline 
ggF &  $H \to \tau \tau$ &{36.1 fb$^{-1}$}\cite{atlasrecent} &  $1.02^{+0.60}_{-0.55}$  \\
VBF&   & &  $1.15^{+0.57}_{-0.53}$  \\
$ttH+tH$&  & &  $1.20^{+1.07}_{-0.93}$  \\
\hline

$ttH+tH$& $ H \to VV $  & {36.1 fb$^{-1}$}\cite{atlasrecent}& $1.72^{+0.56}_{-0.53}$\\
\hline
\end{tabular} 
\end{center}
\caption{Summary of ATLAS Run 2 (24.5-139 fb$^{-1}$(13 TeV)) Higgs signal strength measurements \cite{atlasrecent}. The measurement of VH,$H \rightarrow WW $ is taken from \cite{atlasvhww}.}
\label{tab:atlasrun2}
\end{table}

\end{document}